\documentclass[aps,prl,reprint,groupedaddress]{revtex4-2}

\usepackage{bm}
\usepackage{color}
\usepackage{amssymb}
\usepackage{amsmath}
\usepackage{dcolumn}
\usepackage{colortbl}
\usepackage{graphicx}
\usepackage{hyperref}
\usepackage{mathrsfs}
\usepackage{multirow}
\usepackage{longtable}
\usepackage[table,xcdraw]{xcolor}

\usepackage{url}
\usepackage{ulem}
\usepackage{natbib}
\usepackage{diagbox}
\usepackage{rotating}
\usepackage{enumitem}
\usepackage{booktabs}
\usepackage{threeparttable}
\usepackage[thicklines]{cancel}

\begin{document}

\title{Single- and Double-$\Lambda$ Hypernuclear Correlations Calibrate $\Lambda\Lambda$ Interaction Energies}

\author{Shi Yuan Ding}
\affiliation{MOE Frontiers Science Center for Rare Isotopes, Lanzhou University, Lanzhou 730000, China}
\affiliation{School of Nuclear Science and Technology, Lanzhou University, Lanzhou 730000, China}

\author{Bao Yuan Sun}
\email[Contact author: ]{sunby@lzu.edu.cn}
\affiliation{MOE Frontiers Science Center for Rare Isotopes, Lanzhou University, Lanzhou 730000, China}
\affiliation{School of Nuclear Science and Technology, Lanzhou University, Lanzhou 730000, China}

\begin{abstract}
Double-$\Lambda$ hypernuclei are essential for probing the $\Lambda\Lambda$ interaction in the double-strangeness $S=-2$ sector, yet the scarcity of experimental data severely limits systematic predictions. We present an evaluation framework based on nuclear many-body theory that exploits the intrinsic structural similarity between single-$\Lambda$ and double-$\Lambda$ systems to transfer empirical constraints from the well-mapped $S = -1$ sector to the $S = -2$ sector. By analyzing theoretical deviations of binding energies in light single- and double-$\Lambda$ hypernuclei, we identify a robust linear correlation between two sectors. This correlation enables a statistical evaluation of double-$\Lambda$ separation energies ($B_{\Lambda\Lambda}$) and $\Lambda\Lambda$ interaction energies ($\Delta B_{\Lambda\Lambda}$) for heavier double-$\Lambda$ hypernuclei, by drawing on a wealth of empirical data from the single-$\Lambda$ sector with quantified uncertainties. Our results show that evaluated $\Delta B_{\Lambda\Lambda}$ values, while consistent with existing data, are systematically larger than direct relativistic density functional predictions constrained only by the NAGARA event. This discrepancy suggests that standard mean-field-based extrapolations may underestimate $\Lambda\Lambda$ correlations and other many-body effects, motivating an evaluation-based correction that offers crucial benchmarks for future $S = -2$ experiments at facilities such as HIAF and J-PARC.
\end{abstract}

\maketitle

Hypernuclei provide a unique experimental laboratory for investigating baryon-baryon interactions involving strangeness, including hyperon-nucleon ($YN$) and hyperon-hyperon ($YY$) interactions. These interactions are essential for understanding hypernuclear structure, dense baryonic matter, and the equation of state of neutron stars \cite{Gal2016Rev.Mod.Phys.88.035004, Gibson1995Phys.Rep.257.349, Epelbaum2009Rev.Mod.Phys.81.1773, Hiyama2009Prog.Part.Nucl.Phys.63.339, Lenske2018PPNP98.119, Burgio2021PPNP120.103879, Haidenbauer2026PPNP149.104242}. Among hypernuclear systems, double-$\Lambda$ hypernuclei play a distinctive role, as they provide rare empirical constraints on the in-medium behavior of the $\Lambda\Lambda$ interaction in the double-strangeness sector, $S=-2$, and offer important tests of symmetry-guided baryon-baryon interaction models with strangeness \cite{Fu2022PLB834.137470, Vidana2025EPJA61.59}.

Owing to their low production yields, short lifetimes, and complex weak-decay topologies, only a few double-$\Lambda$ hypernuclear events have been observed over the past several decades \cite{Danysz1963Nucl.Phys.49.121, Prowse1966PRL17.782, Aoki1991PTP85.1287, Aoki2009NPA828.191, Ahn2013PRC88.014003, Ekawa2019PTEP2019.021D02, Miwa2025EPJA61.128}. The NAGARA event, identified as $^{6}_{\Lambda\Lambda}$He, remains the key benchmark, because its clean decay topology enabled a precise determination of the $\Lambda\Lambda$ interaction energy, $\Delta B_{\Lambda\Lambda}=0.67\pm0.17$ MeV \cite{Takahashi2001PRL87.212502, Ahn2013PRC88.014003}. In recent years, next-generation high-intensity beam facilities such as HIAF, together with high-resolution detection techniques, are expected to enhance the production, reconstruction, and identification of double-strangeness hypernuclei \cite{Nakazawa2010NPA835.207, Nakazawa2010PTPS185.335, Acharya2019PLB797.134822, Saito2021Nat.Rev.Phys.3.803, Zhou2022AAPPSBull32.35, He2025NIMA1073.170196, He2025Nat.Commun.16.11084}. Meanwhile, AI-assisted reanalysis of nuclear-emulsion data has opened a new route for double-$\Lambda$ hypernuclear searches and has yielded a $^{13}_{\Lambda\Lambda}$B candidate in a heavier system \cite{He2025Nat.Commun.16.11084}. Nevertheless, existing data remain too limited to quantitatively constrain the $\Lambda\Lambda$ interaction over a broad mass range.

Consequently, unlike the $\Lambda N$ interaction constrained by abundant single-$\Lambda$ hypernuclear data, baryon-baryon interactions in the $S=-2$ sector remain poorly constrained. Since free-space $\Lambda\Lambda$ scattering data are unavailable, present constraints come mainly from a few double-$\Lambda$ hypernuclear events, together with symmetry-based extrapolations, coupled-channel analyses, and lattice-QCD calculations \cite{Vidana2004PRC70.024306, Rijken2010Prog.Theor.Phys.Suppl.185.14, Sasaki2020NPA998.121737, Kamiya2022PRC105.014915, Gal2024PLB857.138973, Haidenbauer2026PPNP149.104242}. These interactions have been employed in few-body and shell model calculations of light double-$\Lambda$ hypernuclei \cite{Hiyama2002PRC66.024007, Hiyama2010PRL104.212502, Garcilazo2013PRL110.012503, Yoshiko2018PRC97.034324, Contessi2019PLB797.134893, Meher2021PRC103.014001}, while density functional descriptions provide a practical framework for systematic studies of heavier systems \cite{Lanskoy1998PRC58.3351, Shen2006PTP115.325, Zhou2007PRC76.034312, Minato2012PRC85.024316, Schulze2013PRC88.024322}. Owing to the sparsity of empirical constraints, predictions for double-$\Lambda$ hypernuclei show appreciable model dependence in regions where direct data are scarce or absent.

When direct empirical information is insufficient, systematic evaluation strategies establish reference values, quantify uncertainties, and separate physical trends from model-dependent effects. Combining limited data with theoretical systematics, correlations, and uncertainty-aware inference, such approaches are widely used in particle physics, astrophysics, cosmology, and nuclear-data evaluation \cite{Ghahramani2015Nature521.7553, Karniadakis2021Nat.Rev.Phys.3.422, Karagiorgi2022Nat.Rev.Phys.4.399, Annala2022PRX12.011058, Brown2018NDS148.1, Boehnlein2022RevModPhys.94.031003}. For double-$\Lambda$ hypernuclei, this perspective motivates an evaluation strategy that combines scarce experimental information with theoretical systematics and correlations, providing additional constraints on quantities that cannot yet be determined directly.

In this Letter, we present an evaluation framework for double-$\Lambda$ hypernuclei assisted by nuclear many-body theory. Using relativistic density functional (RDF) calculations as a realization, we calculate core nuclei and their single- and double-$\Lambda$ partners with available separation-energy data. The resulting theory-experiment deviations exhibit a robust linear correlation between the single- and double-$\Lambda$ binding energies. This correlation reflects the common $\Lambda$-induced core rearrangement in the two systems, while the second $\Lambda$ mainly adds binding without qualitatively changing the core polarization. Exploiting this relation extends the experimental constraints from the $S=-1$ sector to the poorly constrained $S=-2$ sector, yielding evaluated double-$\Lambda$ separation and interaction energies over a broad mass range with quantified uncertainties. The resulting evaluation acts as a correction for missing correlation effects beyond the direct model calculation and provides benchmark references for future double-$\Lambda$ measurements at J-PARC and HIAF.

Building on our previous studies of hypernuclear structure \cite{Ding2022PRC106.054311, Ding2023CPC47.124103, Yang2024PRC110.054320, Ding2025PRC111.014301, Ding2025PRD112.103008}, the present calculations are performed within the RDF framework. The $\Lambda N$ interaction is described by coupling the $\Lambda$ hyperon to the scalar $\sigma$ and vector $\omega$ mesons, with coupling strengths taken from previous relativistic density functional studies constrained by single-$\Lambda$ separation energies \cite{Shen2006PTP115.325, Tu2022APJ925.16, Ding2022PRC106.054311, Ding2023CPC47.124103}. For double-$\Lambda$ hypernuclei, an effective $\Lambda\Lambda$ interaction is included through the hidden-strangeness mesons $\sigma^{*}$ and $\phi$. The $\phi$-$\Lambda$ coupling is fixed by the SU(6) ratio $R_{\phi\Lambda}=g_{\phi\Lambda}/g_{\omega N}=-2\sqrt{2}/3$, while the $\sigma^{*}$-$\Lambda$ coupling is calibrated to the empirical value of the $\Lambda\Lambda$ interaction energy, $\Delta B_{\Lambda\Lambda}$, in $^{6}_{\Lambda\Lambda}$He \cite{Ahn2013PRC88.014003}. This quantity is defined as
\begin{align}
	\Delta B_{\Lambda\Lambda}(^{A+2}_{\Lambda\Lambda}Z)&=B_{\Lambda\Lambda}(^{A+2}_{\Lambda\Lambda}Z)-2B_{\Lambda}(^{A+1}_{\Lambda}Z).\label{eq:IntEnergy}
\end{align}
Here, $B_{\Lambda}$ and $B_{\Lambda\Lambda}$ are the single- and double-$\Lambda$ separation energies, respectively, and $A$ and $Z$ denote the mass and proton numbers of the core nucleus, $^{A}Z$.

Calibration to the NAGARA event alone does not ensure reliable extrapolations to heavier double-$\Lambda$ hypernuclei, where core polarization, density dependence, and correlations may affect the inferred $\Delta B_{\Lambda\Lambda}$. At the same time, progress at J-PARC and the prospect of future double-$\Lambda$ measurements at HIAF call for quantitative reference values to guide candidate selection and data interpretation. To examine whether the more extensive single-$\Lambda$ hypernuclear data can constrain double-$\Lambda$ observables, we perform a correlation analysis based on RDF calculations across core nuclei and their hypernuclear counterparts. The analyzed quantities include the separation energies $B_{\Lambda}$ and $B_{\Lambda\Lambda}$, the interaction energy $\Delta B_{\Lambda\Lambda}$, as well as theory-experiment deviations in separation energies, $\delta B_{\Lambda}$ and $\delta B_{\Lambda\Lambda}$, and binding energies, $\delta E_{B}(^{A}Z)$, $\delta E_{B}(^{A+1}_{\Lambda}Z)$, and $\delta E_{B}(^{A+2}_{\Lambda\Lambda}Z)$. These quantities are used to identify correlations relevant for evaluating double-$\Lambda$ observables, allowing the resulting evaluation to compensate for missing correlation effects in direct RDF predictions. Details of the single-$\Lambda$ hypernuclear calculations are given in Refs. \cite{Ding2022PRC106.054311, Ding2023CPC47.124103}.
\begin{figure}[htbp]
	\centering
	\includegraphics[width=0.48\textwidth]{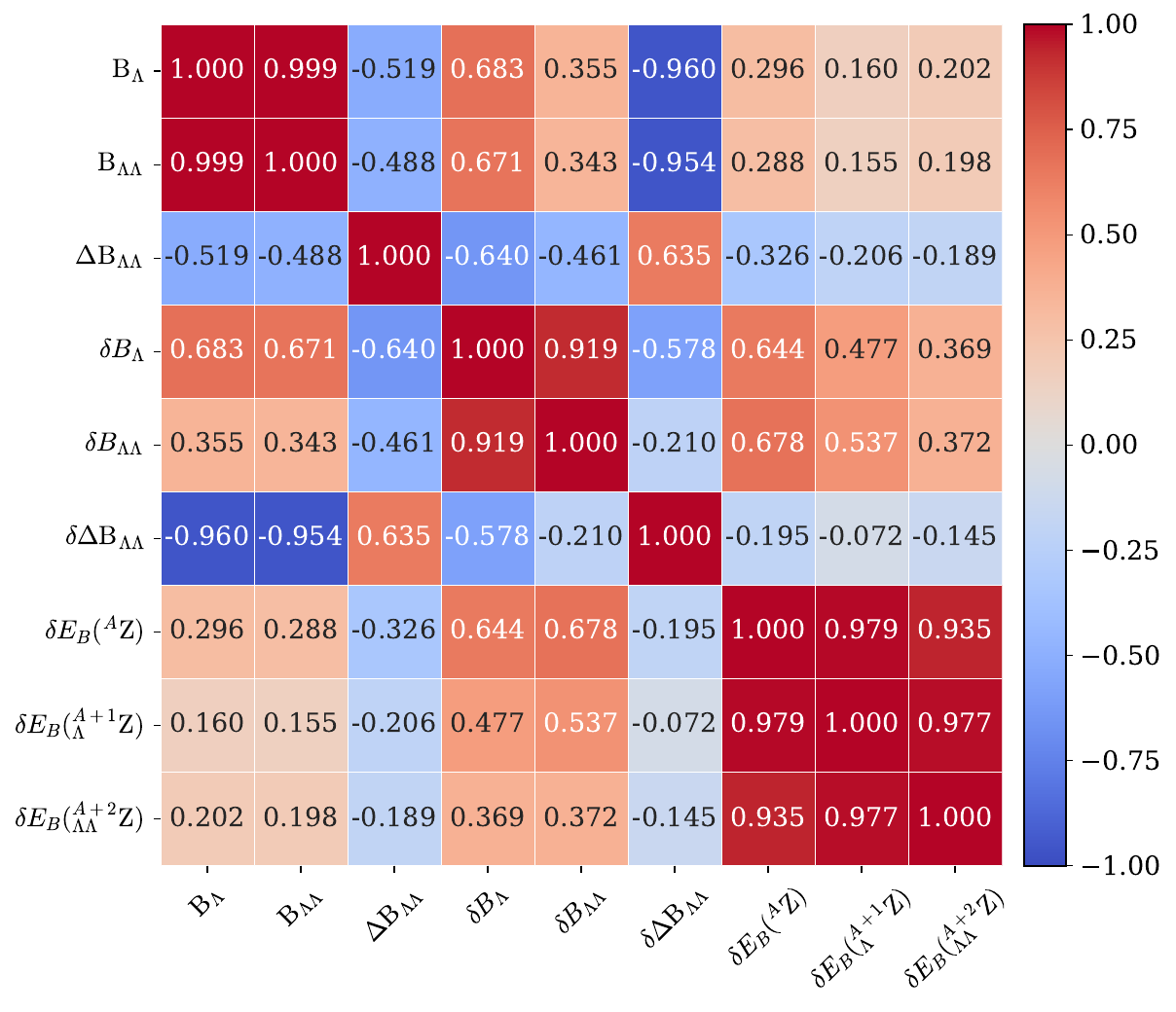}
	\caption{A Pearson correlation matrix is constructed for various hypernuclear quantities obtained from RDF models, including the hyperon separation energies $B_{\Lambda}$ and $B_{\Lambda\Lambda}$, the $\Lambda\Lambda$ interaction energy $\Delta B_{\Lambda\Lambda}$, and their corresponding deviations from experimental data, defined in Eq. \eqref{eq:LambdaLambdaEB}, namely $\delta B_{\Lambda}$, $\delta B_{\Lambda\Lambda}$, and $\delta \Delta B_{\Lambda\Lambda}$. The statistical analysis involves the core nuclei $^{4}$He, $^{8}$Be, $^{9}$Be, and $^{11}$B, along with their corresponding single-$\Lambda$ and double-$\Lambda$ hypernuclear partners. It also includes the deviations in binding energies for the core nuclei, single-$\Lambda$, and double-$\Lambda$ hypernuclei, denoted by $\delta E_{B}(^{A}Z)$, $\delta E_{B}(^{A+1}_{\Lambda}Z)$, and $\delta E_{B}(^{A+2}_{\Lambda\Lambda}Z)$, respectively.}\label{Fig:LinearCorrelationMatrix}
\end{figure}

It should be emphasized that hypernuclear binding energies are not directly measured, but reconstructed from the binding energy of the core nucleus and the hyperon separation energies. Thus, the quoted absolute binding energies may depend on the adopted values of $E_{B}(^{A}Z)$, $B_{\Lambda}$, and $B_{\Lambda\Lambda}$. For a double-$\Lambda$ hypernucleus, we use
\begin{align}
	E_{B}(^{A+2}_{\Lambda\Lambda}Z)&=E_{B}(^{A}Z)-B_{\Lambda\Lambda}(^{A+2}_{\Lambda\Lambda}Z). \label{eq:DeltaLambdaLambda}
\end{align}
The empirical core binding energies are taken from Ref. \cite{Wang2021CPC45.030003}, and the experimental hyperon separation energies from Refs. \cite{Ahn2013PRC88.014003, Ekawa2019PTEP2019.021D02, HypernucleiWeb2022, He2025Nat.Commun.16.11084}.

\begin{figure}[htbp]
	\centering
	\includegraphics[width=0.48\textwidth]{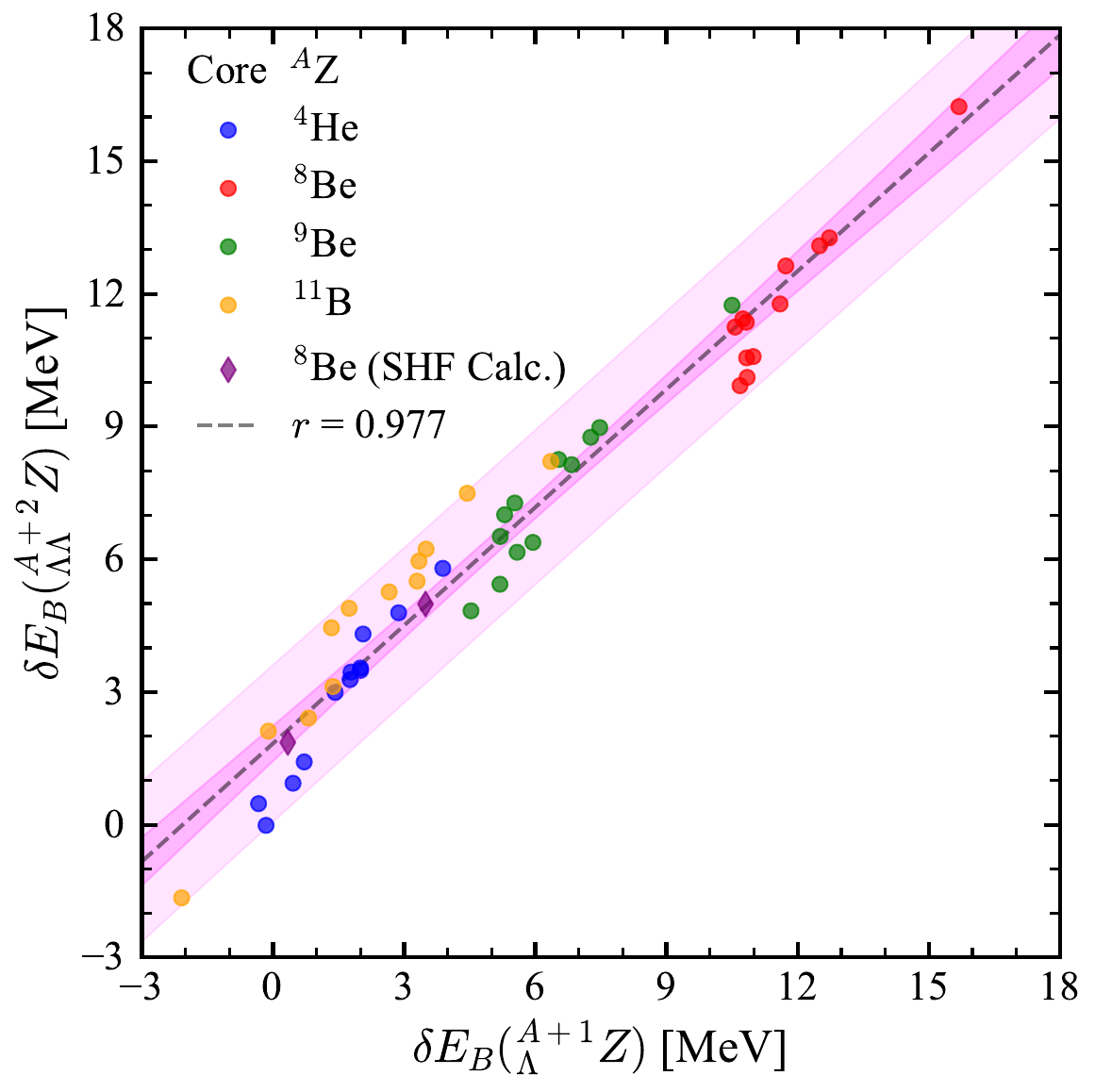}
	\caption{The black dashed lines show linear fits to the RDF differences between theoretical and experimental binding energies of single-$\Lambda$ and double-$\Lambda$ hypernuclei built on $^{4}$He, $^{8}$Be, $^{9}$Be, and $^{11}$B cores, corresponding to $\delta E_{B}(^{A+2}_{\Lambda\Lambda}Z)=0.890~\delta E_{B}(^{A+1}_{\Lambda}Z)+1.837$. The pink and light-pink shaded regions indicate the 95\% confidence and prediction intervals, respectively. As a reference, Skyrme-Hartree-Fock results for $^{8}$Be from Ref.~\cite{Zhou2007PRC76.034312} are also shown as purple diamonds.}\label{Fig:LinearCorrelation}
\end{figure}
Based on these inputs, we analyze the four double-$\Lambda$ hypernuclei with available experimental information, corresponding to the core nuclei $^{4}$He, $^{8}$Be, $^{9}$Be, and $^{11}$B. The data for the first three systems are taken from Refs. \cite{Ahn2013PRC88.014003, Ekawa2019PTEP2019.021D02}, while those for the double-$\Lambda$ hypernucleus with the $^{11}$B core are from the recent AI-assisted emulsion analysis of Ref. \cite{He2025Nat.Commun.16.11084}. From the corresponding RDF results, we construct the Pearson correlation matrix among the separation energies and the theory-experiment deviations in separation and binding energies, as shown in Fig.~\ref{Fig:LinearCorrelationMatrix}. This matrix identifies the physical quantities most strongly correlated with double-$\Lambda$ properties and thus most effective in constraining them. These correlations provide a basis for weakly model-dependent extrapolations of the binding and separation energies of unknown double-$\Lambda$ hypernuclei within the considered RDF sets.

As shown in Fig.~\ref{Fig:LinearCorrelationMatrix}, several strong correlations emerge among the quantities considered here. We focus on correlations involving theory-experiment deviations, which incorporate available experimental constraints into the evaluation of double-$\Lambda$ observables. In particular, the binding-energy deviations of single- and double-$\Lambda$ hypernuclei, $\delta E_{B}(^{A+1}_{\Lambda}Z)$ and $\delta E_{B}(^{A+2}_{\Lambda\Lambda}Z)$, show a strong linear correlation with a Pearson coefficient of $0.977$. This correlation links the systematic deviations in the single- and double-$\Lambda$ sectors and provides the basis for evaluating double-$\Lambda$ interaction energies. For any observable $\mathcal{O}$, the deviation is defined as
\begin{align}
	\delta \mathcal{O}=\mathcal{O}^{\rm{RDF}}-\mathcal{O}^{\rm{Expt}}\label{eq:LambdaLambdaEB}
\end{align}
where the superscripts ``RDF" and ``Expt" denote RDF predictions and experimental values, respectively.

Guided by the correlation identified above, we examine the relation between $\delta E_{B}(^{A+1}_{\Lambda}Z)$ and $\delta E_{B}(^{A+2}_{\Lambda\Lambda}Z)$ in Fig.~\ref{Fig:LinearCorrelation}. A linear fit to the results gives
\begin{align}
	\delta E_{B}(^{A+2}_{\Lambda\Lambda}Z)=0.890~\delta E_{B}(^{A+1}_{\Lambda}Z)+1.837 \label{eq:LinearCorrelation}
\end{align}
From the average half-width of the 95\% confidence band, shown in pink in Fig.~\ref{Fig:LinearCorrelation}, we assign a characteristic evaluation uncertainty of $\sigma_{\rm Eval}=0.422$~MeV to the fitted correlation and include it in the subsequent evaluation of double-$\Lambda$ observables. As a cross-check, we also compare with the Skyrme-Hartree-Fock results of Ref.~\cite{Zhou2007PRC76.034312} for the $^{8}$Be-core system, which are marked by purple diamonds.

It should be noted that the binding-energy deviation is not a directly observable quantity, but is inferred relative to a theoretical reference. Therefore, quantitative discrepancies between model predictions and values extracted from experimental information are not unexpected. These discrepancies may indicate the difficulty of achieving a unified self-consistent description of different hypernuclear structures, since missing correlations can make a model that reproduces single-$\Lambda$ hypernuclei less reliable for double-$\Lambda$ systems. Such discrepancies, however, do not obscure the characteristic correlation between the single- and double-$\Lambda$ sectors, which is supported in the light-mass region. Applying the same analysis within other theoretical frameworks, together with future data on heavier double-$\Lambda$ hypernuclei, would provide important tests of the robustness, model dependence, and possible mass dependence of this correlation.

The established linear correlation is used as an evaluation-based correction to estimate the binding-energy deviations of double-$\Lambda$ systems for target cores beyond the four light cores included in the correlation analysis, namely $^{4}$He, $^{8}$Be, $^{9}$Be, and $^{11}$B. For each RDF model, the single-$\Lambda$ deviations $\delta E_{B}(^{A+1}_{\Lambda}Z)$ are obtained from hypernuclei with available experimental separation energies and then inserted into Eq.~\eqref{eq:LinearCorrelation} to determine the corresponding double-$\Lambda$ deviations, $\delta E_{B}^{\rm Eval}(^{A+2}_{\Lambda\Lambda}Z)$. These evaluated deviations, with the fitting uncertainty $\sigma_{\rm Eval}$ propagated through the linear correlation, are used to correct the RDF predictions. This provides a systematic model-by-model propagation of experimental constraints from single-$\Lambda$ to double-$\Lambda$ hypernuclei.

Combining the evaluated deviations $\delta E^{\rm Eval}_{B}(^{A+2}_{\Lambda\Lambda}Z)$ with Eqs.~\eqref{eq:DeltaLambdaLambda} and \eqref{eq:LambdaLambdaEB}, we obtain
\begin{align}
	B^{\rm Eval}_{\Lambda\Lambda}=\delta E^{\rm Eval}_{B}(^{A+2}_{\Lambda\Lambda}Z)-E^{\rm RDF}_{B}(^{A+2}_{\Lambda\Lambda}Z)+E^{\rm Expt}_{B}(^{A}Z).\label{eq:BLambdaLambdaExp}
\end{align}
Eq.~\eqref{eq:BLambdaLambdaExp} gives the evaluated double-$\Lambda$ separation energy $B^{\rm Eval}_{\Lambda\Lambda}$ for each RDF model, where $E^{\rm RDF}_{B}(^{A+2}_{\Lambda\Lambda}Z)$ is the corresponding theoretical binding energy and $E^{\rm Expt}_{B}(^{A}Z)$ is the experimental core-nucleus binding energy taken from Ref.~\cite{Wang2021CPC45.030003}. The uncertainty of $B^{\rm Eval}_{\Lambda\Lambda}$ is assigned as $\sigma_{\rm Eval}$, originating from the evaluated deviation $\delta E^{\rm Eval}_{B}(^{A+2}_{\Lambda\Lambda}Z)$ propagated through the linear correlation. For a given RDF model, the second term in Eq.~\eqref{eq:BLambdaLambdaExp} is fixed, while the uncertainty of the core binding energy in the third term is negligible in the present analysis.

To further reduce the model dependence among RDF descriptions, we select the functionals used in the final evaluation by requiring consistency with the available double-$\Lambda$ data. For each RDF functional, the correlation-based procedure gives an evaluated separation energy $B_{\Lambda\Lambda}^{\rm Eval}$ with its uncertainty. A functional is retained only if this evaluated interval overlaps with the corresponding experimental interval for all four measured double-$\Lambda$ hypernuclei, with core nuclei $^{4}$He, $^{8}$Be, $^{9}$Be, and $^{11}$B, using the experimental values compiled in Refs.~\cite{Ahn2013PRC88.014003, Ekawa2019PTEP2019.021D02, He2025Nat.Commun.16.11084}.

Based on the retained RDF models, we perform an overall evaluation of the double-$\Lambda$ separation energies, obtaining values and uncertainty ranges that do not rely on any single functional. For each double-$\Lambda$ hypernucleus, the central value of $B_{\Lambda\Lambda}^{\rm Eval}$ is taken as the average over the retained RDF models, while the model-to-model spread is denoted by $\sigma_{\rm RDF}$. In addition, each evaluated value carries the uncertainty $\sigma_{\rm Eval}$ from the linear-correlation analysis. The final uncertainty $\sigma_{B_{\Lambda\Lambda}^{\rm Eval}}$ is therefore taken as $\sqrt{\sigma_{\rm RDF}^{2}+\sigma_{\rm Eval}^{2}}$. The resulting evaluated double-$\Lambda$ separation energies, along with the unevaluated direct RDF predictions $B_{\Lambda\Lambda}^{\rm RDF}$ and their spreads, are listed in Table~\ref{Tab:DoubleLambdaSeparationEnergy-RDF}.
\begin{table*}[hbpt]
	\centering
	\caption{Separation energies $B_{\Lambda\Lambda}$, in MeV, for double-$\Lambda$ hypernuclei obtained from RDF models. Here, $B_{\Lambda\Lambda}^{\rm Eval}$ denotes the mean values and corresponding uncertainties obtained from the retained RDF models after the evaluation procedure, while $B_{\Lambda\Lambda}^{\rm RDF}$ denotes those obtained directly from the RDF model calculations calibrated only to the NAGARA event.}
	\label{Tab:DoubleLambdaSeparationEnergy-RDF}
	\renewcommand{\arraystretch}{2.0}
	\doublerulesep 0.1pt
	\tabcolsep 2.3pt
	\begin{tabular}{cccccccccc}
		\hline\hline
		& $^{6}_{\Lambda\Lambda}$He  
		& $^{8}_{\Lambda\Lambda}$Li  
		& $^{9}_{\Lambda\Lambda}$Li  
		& $^{10}_{\Lambda\Lambda}$Be 
		& $^{11}_{\Lambda\Lambda}$Be 
		& $^{12}_{\Lambda\Lambda}$B
		& $^{13}_{\Lambda\Lambda}$B  
		& $^{13}_{\Lambda\Lambda}$C  
		& $^{14}_{\Lambda\Lambda}$C  \\
		\hline
		$B_{\Lambda\Lambda}^{\rm{Eval}}$ 
		& $6.895^{+0.478}_{-0.478}$  
		& $11.915^{+0.481}_{-0.481}$ 
		& $14.453^{+0.465}_{-0.465}$ 
		& $14.949^{+0.454}_{-0.454}$ 
		& $18.647^{+0.455}_{-0.455}$ 
		& $21.727^{+0.484}_{-0.484}$
		& $24.494^{+0.583}_{-0.583}$ 
		& $24.401^{+0.559}_{-0.559}$ 
		& $26.047^{+0.692}_{-0.692}$  \\
		$B_{\Lambda\Lambda}^{\rm{RDF}}$ 
		& $3.679^{+0.315}_{-0.315}$  
		& $9.411^{+0.300}_{-0.300}$  
		& $12.387^{+0.203}_{-0.203}$ 
		& $15.132^{+0.125}_{-0.125}$ 
		& $17.505^{+0.215}_{-0.215}$ 
		& $19.917^{+0.394}_{-0.394}$
		& $22.660^{+0.718}_{-0.718}$ 
		& $22.554^{+0.676}_{-0.676}$ 
		& $25.234^{+0.983}_{-0.983}$  \\
		\hline\hline
		& $^{17}_{\Lambda\Lambda}$N  
		& $^{17}_{\Lambda\Lambda}$O  
		& $^{29}_{\Lambda\Lambda}$Si
		& $^{33}_{\Lambda\Lambda}$S  
		& $^{41}_{\Lambda\Lambda}$Ca 
		& $^{52}_{\Lambda\Lambda}$V  
		& $^{90}_{\Lambda\Lambda}$Y  
		& $^{140}_{\Lambda\Lambda}$La 
		& $^{209}_{\Lambda\Lambda}$Pb \\
		\hline
		$B_{\Lambda\Lambda}^{\rm{Eval}}$ 
		& $28.271^{+0.532}_{-0.532}$ 
		& $27.567^{+0.513}_{-0.513}$ 
		& $36.540^{+0.688}_{-0.688}$
		& $38.549^{+0.748}_{-0.748}$ 
		& $39.929^{+0.581}_{-0.581}$ 
		& $42.843^{+0.746}_{-0.746}$ 
		& $48.597^{+0.986}_{-0.986}$ 
		& $50.901^{+1.292}_{-1.292}$ 
		& $54.866^{+1.657}_{-1.657}$  \\
		$B_{\Lambda\Lambda}^{\rm{RDF}}$ 
		& $25.213^{+0.603}_{-0.603}$ 
		& $25.174^{+0.574}_{-0.574}$ 
		& $36.905^{+0.753}_{-0.753}$
		& $38.975^{+1.012}_{-1.012}$ 
		& $38.590^{+0.596}_{-0.596}$ 
		& $42.749^{+0.543}_{-0.543}$ 
		& $47.479^{+0.486}_{-0.486}$ 
		& $50.031^{+0.556}_{-0.556}$ 
		& $52.637^{+0.702}_{-0.702}$  \\
		\hline\hline
	\end{tabular}
\end{table*}

As shown in Table~\ref{Tab:DoubleLambdaSeparationEnergy-RDF}, the unevaluated RDF predictions for $B_{\Lambda\Lambda}$ are generally smaller than the evaluated values, particularly for light and some medium-mass nuclei. Given the good agreement of the evaluated values with the available data, this systematic difference suggests that mean-field calculations calibrated only to the NAGARA event underestimate the binding associated with double-$\Lambda$ dynamics in real hypernuclei. Extended theoretical studies have shown that few-body and cluster correlations, core polarization, and medium modifications of the effective $\Lambda\Lambda$ interaction can affect double-$\Lambda$ hypernuclear binding energies~\cite{Hiyama2002PRC66.024007, Yoshiko2018PRC97.034324, Schulze2013PRC88.024322}. By contrast, the present evaluation propagates the better-constrained empirical information from the single-$\Lambda$ sector through the established linear correlation, thereby providing a phenomenological estimate of effects not explicitly included at the current mean-field level.

\begin{figure}[htbp]
	\centering
	\includegraphics[width=0.48\textwidth]{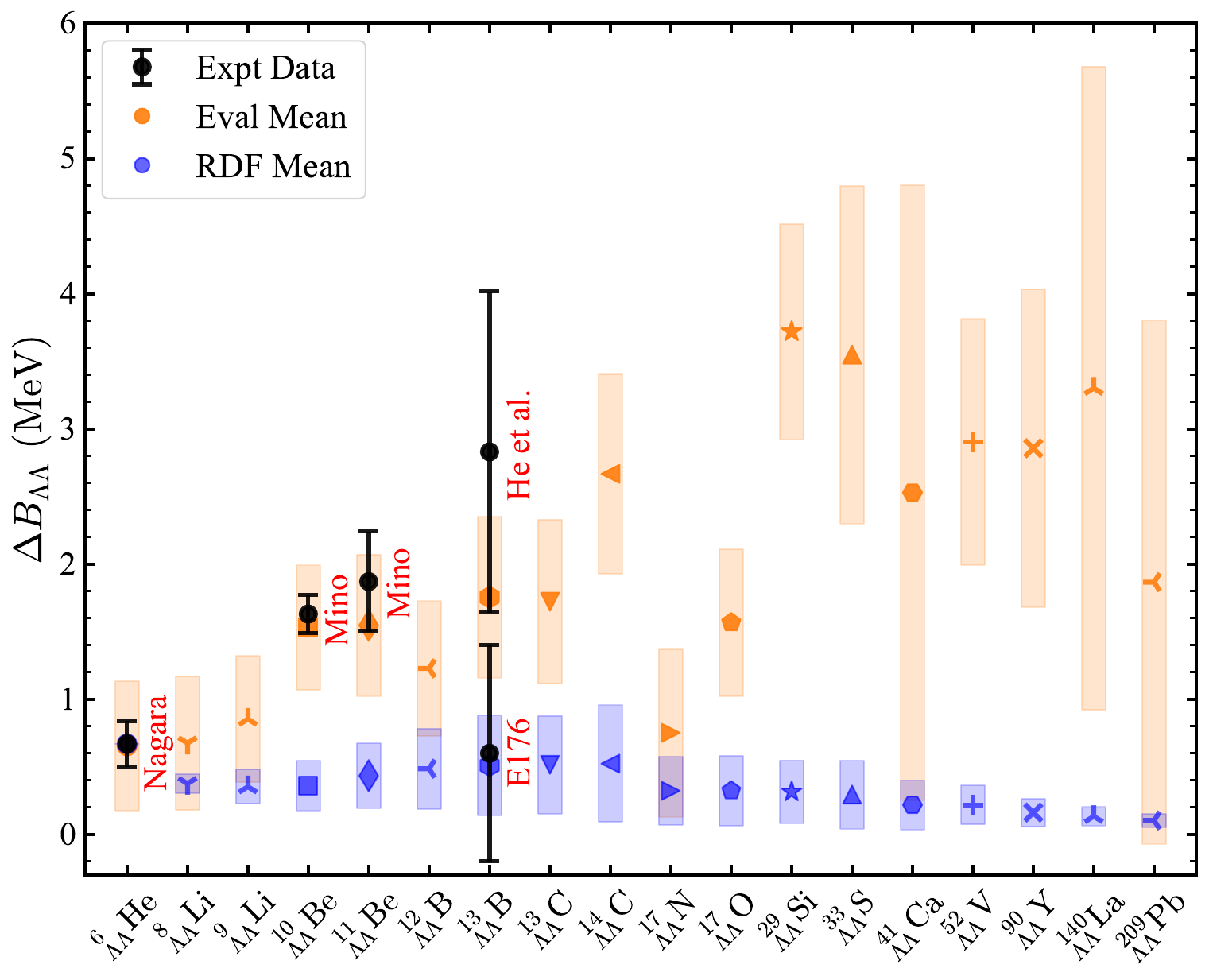}
	\caption{The $\Lambda\Lambda$ interaction energies of double-$\Lambda$ hypernuclei are calculated with a set of RDF models. Blue symbols denote the mean values predicted by the models, and orange symbols represent the values evaluated from the linear relation in Eq.~(\ref{eq:LinearCorrelation}). The shaded bands indicate the corresponding $1\sigma$ uncertainties. Available experimental data are included for comparison, with error bars taken from Refs.~\cite{Aoki2009NPA828.191, Ahn2013PRC88.014003, Ekawa2019PTEP2019.021D02, He2025Nat.Commun.16.11084}.}\label{Fig:IntEnergy}
\end{figure}

With the evaluated double-$\Lambda$ separation energies, we extract the corresponding $\Lambda\Lambda$ interaction energies according to Eq.~\eqref{eq:IntEnergy}, propagating also the experimental uncertainty of the single-$\Lambda$ separation energy, $\sigma_{B_{\Lambda}^{\rm Expt}}$. Assuming $\sigma_{B_{\Lambda}^{\rm Expt}}$ is independent of $\sigma_{B_{\Lambda\Lambda}^{\rm Eval}}$, the uncertainty of the evaluated interaction energy is
\begin{align}
	\sigma_{\Delta B_{\Lambda\Lambda}^{\rm Eval}}=\sqrt{\sigma_{B_{\Lambda\Lambda}^{\rm Eval}}^{2}+4\sigma_{B_{\Lambda}^{\rm Expt}}^{2}}.
\end{align}
The resulting evaluated interaction energies are shown in Fig.~\ref{Fig:IntEnergy}. The orange symbols denote the evaluated means, the blue symbols represent the RDF means without correlation-based evaluation, and the shaded bands indicate the corresponding uncertainties. Available experimental data are shown by black error bars; for the $^{11}$B-core system, the earlier E176 result is also included for comparison~\cite{Aoki2009NPA828.191}.

As shown in Fig.~\ref{Fig:IntEnergy}, the evaluated $\Delta B_{\Lambda\Lambda}$ are generally larger than the unevaluated RDF predictions but reproduce the existing double-$\Lambda$ data within uncertainties. All values stay below $5$~MeV, consistent with previous studies~\cite{Acharya2019PLB797.134822,Miwa2025EPJA61.128,He2025Nat.Commun.16.11084}, and show a characteristic mass dependence, with larger values in the medium-mass region. This suggests that RDF calculations calibrated only to the NAGARA event may underestimate $\Lambda\Lambda$ binding in some systems, while the present data-driven evaluation partly accounts for effects beyond the mean-field description. In light nuclei, few-body cluster correlations and core polarization can strongly affect the extracted $\Lambda\Lambda$ binding~\cite{Hiyama2010PRL104.212502,Yoshiko2018PRC97.034324}, whereas in medium-mass nuclei, increased spatial overlap of $\Lambda$ wave functions enhances sensitivity to core polarization and effective $\Lambda\Lambda$ correlations. In heavier systems, the single-particle binding of $\Lambda$ hyperons saturates, so the additional binding from two $\Lambda$ particles grows more slowly with mass.

The uncertainty of $\Delta B_{\Lambda\Lambda}^{\rm Eval}$ arises not only from the spread among the retained RDF models and the linear-correlation analysis, but also from the experimental uncertainty of the single-$\Lambda$ separation energies. This contribution is particularly significant in medium- and heavy-mass nuclei, where single-$\Lambda$ data are often less precise. Improved and systematic measurements of single-$\Lambda$ hypernuclei, especially in the medium-mass region, would reduce the uncertainty of the extracted interaction energies. In addition, future measurements of double-$\Lambda$ hypernuclei across a wider mass range will provide stringent tests of the present evaluation scheme, the mass dependence of $\Delta B_{\Lambda\Lambda}$, and the density dependence of effective strange-baryon interactions in RDF models.

To assess the sensitivity and robustness of our results, we repeated the linear fitting excluding $^{10}_{\Lambda\Lambda}$Be, which may be affected by deformation and experimental ambiguities~\cite{Hiyama2002PRC66.024007}. This slightly increases the evaluated mean for the $^{8}$Be-core system and modestly reduces $\Delta B_{\Lambda\Lambda}^{\rm Eval}$ in the heavy-mass region, while leaving the mass dependence and results for other nuclei essentially unchanged. The stability of this trend supports the robustness of the correlation-based evaluation, providing quantified constraints on interaction energies where direct double-$\Lambda$ data are scarce. These evaluated systematics offer microscopic input for future studies of double- and multi-strange hypernuclei near the drip line, where weak binding, continuum effects, and hyperon-hyperon correlations, including possible $\Lambda\Lambda$ pairing, may affect shell evolution and stability~\cite{Margueron2017PRC96.054317,Rong2020PLB807.135533,Guo2022PRC105.034322}. In dense matter, the in-medium behavior of the $\Lambda\Lambda$ interaction and associated pairing gap can influence the composition, neutrino emissivity, and cooling of neutron stars once hyperons appear~\cite{Schaab1998APJ504.L99,Tanigawa2003PRC68.015801,Guven2018PRC98.014318}. Thus, reliable constraints on interaction energies link hypernuclear observables to the microscopic physics of hyperon-rich compact stars.

In summary, we studied double-$\Lambda$ hypernuclei within a theory-assisted evaluation framework, realized by RDF calculations with NAGARA-constrained $\Lambda\Lambda$ interactions. We identify a linear correlation between theory-experiment deviations in single- and double-$\Lambda$ binding energies. Exploiting this correlation, we propagate empirical constraints from the well-measured single-$\Lambda$ sector to double-$\Lambda$ systems, obtaining evaluated $B_{\Lambda\Lambda}$ and $\Delta B_{\Lambda\Lambda}$ with quantified uncertainties. The evaluated $\Delta B_{\Lambda\Lambda}$ remain below $5$~MeV, peak in medium-mass nuclei, and are generally larger than unevaluated RDF predictions, indicating that NAGARA-only calibration may underestimate structure- and medium-dependent correlations.

This evaluation-based correction thus compensates for missing correlation effects in direct RDF predictions by absorbing physics beyond the model space, such as clustering, core polarization, coupled-channel $\Lambda\Lambda$-$\Xi N$ mixing, and possible $\Lambda\Lambda$ pairing. Although these microscopic mechanisms are not disentangled here, the resulting benchmarks provide a crucial reference for future experiments at HIAF and J-PARC and critical input for neutron-star matter studies. More broadly, the correlation-based evaluation framework offers a generalizable methodology for statistical inference in data-sparse systems across nuclear and particle physics.

\begin{acknowledgments}
This work was partly supported by the National Natural Science Foundation of China (No. 11875152 and No. 12275111).
\end{acknowledgments}

\newpage
\onecolumngrid
\begin{center}
	\textbf{Appendix}
\end{center}
\twocolumngrid

\section{A. Relativistic density functionals and hyperon couplings}\label{sec:interactions}

The nucleon interactions are described by well-established nonlinear RMF functionals such as PK1, TM1, and NL-SH, together with six density-dependent RMF functionals, namely TW99, PKDD, DD-LZ1, DD-ME2, DD-ME$\delta$, and DD-MEX. In addition, we employ three density-dependent RHF functionals, PKO1, PKO2, and PKO3. The calculations are performed in the spherical relativistic density functional framework. The Dirac equations for baryons are solved in a radial box size of $R=20$ fm with a step of $0.1$ fm.

Pairing correlations in open-shell systems are treated within the BCS method using the Gogny force D1S in the $nn$ and $pp$ pairing channels. In odd systems, the blocking effect is included for the last valence nucleon or hyperon. The ground-state configuration is determined by comparing the total binding energies obtained with different blocked single-particle orbits near the Fermi surface.

The $\Lambda N$ interaction is introduced through the coupling of the $\Lambda$ hyperon to the scalar $\sigma$ and vector $\omega$ mesons. The vector coupling ratio is fixed at $R_{\omega\Lambda}=g_{\omega\Lambda}/g_{\omega N}=0.666$, while the scalar ratios $R_{\sigma\Lambda}$ are taken from previous single-$\Lambda$ RDF studies constrained by experimental separation energies \cite{Shen2006PTP115.325, Tu2022APJ925.16, Ding2022PRC106.054311, Ding2023CPC47.124103}. The effective $\Lambda\Lambda$ interaction is described by the hidden-strangeness mesons $\sigma^{*}$ and $\phi$. The vector coupling is fixed by the SU(6) relation $R_{\phi\Lambda}=-2\sqrt{2}/3$,
whereas the scalar coupling $R_{\sigma^{*}\Lambda}$ is determined separately for each RDF functional by reproducing the NAGARA value $\Delta B_{\Lambda\Lambda}(^{6}_{\Lambda\Lambda}{\rm He})=0.67$~MeV \cite{Ahn2013PRC88.014003}. The adopted coupling ratios are listed in Table~\ref{tab:couplings}.
\begin{table}[hbpt]
	\centering
	\caption{Hyperon coupling ratios used in the RDF calculations. The $\Lambda N$ scalar ratios $R_{\sigma\Lambda}=g_{\sigma\Lambda}/g_{\sigma N}$ are taken from single-$\Lambda$ RDF fits to experimental separation energies~\cite{Shen2006PTP115.325, Tu2022APJ925.16, Ding2022PRC106.054311, Ding2023CPC47.124103}. The vector couplings are fixed as $R_{\omega\Lambda}=g_{\omega\Lambda}/g_{\omega N}=0.666$ and $R_{\phi\Lambda}=-2\sqrt{2}/3$. The $\Lambda\Lambda$ scalar ratios $R_{\sigma^{*}\Lambda}=g_{\sigma^{*}\Lambda}/g_{\sigma N}$ are fitted to $\Delta B_{\Lambda\Lambda}(^{6}_{\Lambda\Lambda}{\rm He})=0.67$~MeV~\cite{Ahn2013PRC88.014003}.}
	\label{tab:couplings}
	\renewcommand{\arraystretch}{1.35}
	\doublerulesep 0.1pt \tabcolsep 26.0pt
	\begin{tabular}{lcc}
			\hline\hline
			& $R_{\sigma\Lambda}$ & $R_{\sigma^{*}\Lambda}$ \\ \hline
			PK1           & 0.618               & 0.549                      \\
			TM1           & 0.621               & 0.548                      \\
			NL-SH         & 0.621               & 0.540                      \\
			PKO1          & 0.596               & 0.571                      \\
			PKO2          & 0.591               & 0.549                      \\
			PKO3          & 0.594               & 0.519                      \\
			TW99          & 0.617               & 0.530                      \\
			PKDD          & 0.620               & 0.529                      \\
			DD-LZ1        & 0.615               & 0.517                      \\
			DD-ME2        & 0.620               & 0.534                      \\
			DD-ME$\delta$ & 0.625               & 0.513                      \\
			DD-MEX        & 0.618               & 0.537                      \\ 
			\hline\hline
		\end{tabular}
\end{table}

\section{B. Evaluation of double-$\Lambda$ binding energy deviations}\label{sec:evaluation}
\begin{sidewaystable}
	\centering
	\caption{Evaluated double-$\Lambda$ binding-energy deviations $\delta E_{B}^{\rm Eval}(^{A+2}_{\Lambda\Lambda}Z)$ in MeV, obtained from Eq.~\eqref{eq:LinearCorrelation}. For each RDF functional, the associated fitting uncertainty is $\sigma_{\rm Eval}=0.422$~MeV.}
	\label{tab:delta_eval}
	\renewcommand{\arraystretch}{1.05}
	\setlength{\tabcolsep}{11.0pt}
	\begin{tabular}{rrrrrrrrrrrrr}
		\hline\hline
		& PK1    & TM1    & NL-SH  & PKO1   & PKO2   & PKO3   & TW99   & PKDD   & DD-LZ1 & DD-ME2 & DD-ME$\delta$ & DD-MEX  \\
		$^{6}_{\Lambda\Lambda}\mathrm{He}$   & 2.472  & 1.696  & 2.244  & 3.102  & 3.617  & 3.670  & 5.293  & 3.406  & 4.394  & 3.620  & 1.544         & 3.424   \\
		$^{8}_{\Lambda\Lambda}\mathrm{Li}$   & 4.325  & 3.759  & 4.209  & 4.289  & 4.652  & 5.057  & 7.782  & 5.521  & 5.109  & 5.293  & 4.017         & 4.540   \\
		$^{9}_{\Lambda\Lambda}\mathrm{Li}$   & 5.128  & 4.757  & 4.937  & 5.102  & 5.315  & 5.990  & 9.052  & 6.541  & 5.783  & 6.368  & 5.240         & 5.424   \\
		$^{10}_{\Lambda\Lambda}\mathrm{Be}$  & 11.610 & 11.487 & 11.341 & 11.398 & 11.241 & 12.274 & 15.797 & 13.161 & 11.476 & 12.961 & 12.157        & 11.470  \\
		$^{11}_{\Lambda\Lambda}\mathrm{Be}$  & 6.806  & 7.127  & 6.456  & 6.758  & 6.551  & 7.650  & 11.180 & 8.487  & 5.867  & 8.303  & 7.914         & 6.462   \\
		$^{12}_{\Lambda\Lambda}\mathrm{B}$   & 5.297  & 6.203  & 4.899  & 5.185  & 4.829  & 6.057  & 9.791  & 7.081  & 3.331  & 6.890  & 7.233         & 4.378   \\
		$^{13}_{\Lambda\Lambda}\mathrm{B}$   & 3.057  & 4.771  & 2.561  & 3.387  & 3.027  & 4.203  & 7.491  & 4.953  & -0.022 & 4.808  & 5.790         & 1.743   \\
		$^{13}_{\Lambda\Lambda}\mathrm{C}$   & 3.302  & 4.914  & 2.778  & 3.128  & 2.789  & 3.980  & 7.802  & 5.196  & 0.390  & 5.063  & 6.023         & 2.034   \\
		$^{14}_{\Lambda\Lambda}\mathrm{C}$   & 3.518  & 6.242  & 3.031  & 3.985  & 3.674  & 4.769  & 7.959  & 5.564  & -0.772 & 5.530  & 7.524         & 1.644   \\
		$^{17}_{\Lambda\Lambda}\mathrm{N}$   & 1.659  & 0.849  & 1.797  & 1.985  & 2.980  & 2.366  & 5.817  & 2.453  & 1.094  & 2.347  & 0.208         & -0.025  \\
		$^{17}_{\Lambda\Lambda}\mathrm{O}$   & 1.149  & 0.295  & 1.244  & 0.987  & 2.011  & 1.390  & 5.417  & 1.997  & 0.619  & 1.880  & -0.208        & -0.497  \\
		$^{29}_{\Lambda\Lambda}\mathrm{Si}$  & 4.163  & 6.296  & 2.897  & 3.673  & 2.959  & 5.088  & 9.615  & 5.072  & 1.477  & 5.903  & 7.831         & -0.872  \\
		$^{33}_{\Lambda\Lambda}\mathrm{S}$   & 6.275  & 8.224  & 6.268  & 5.207  & 4.790  & 6.262  & 10.377 & 6.293  & -0.357 & 5.996  & 8.580         & -1.297  \\
		$^{41}_{\Lambda\Lambda}\mathrm{Ca}$  & 1.823  & -0.058 & 3.203  & 0.676  & 2.786  & 0.626  & 8.757  & 2.174  & -0.026 & 1.647  & -1.096        & -4.384  \\
		$^{52}_{\Lambda\Lambda}\mathrm{V}$   & 5.298  & 5.586  & 4.425  & 4.785  & 5.148  & 5.416  & 12.911 & 5.508  & 5.118  & 6.786  & 6.293         & -2.473  \\
		$^{90}_{\Lambda\Lambda}\mathrm{Y}$   & 1.331  & -0.029 & 0.906  & 0.375  & 2.357  & 0.035  & 12.356 & 1.220  & 1.921  & 2.514  & 0.839         & -11.147 \\
		$^{140}_{\Lambda\Lambda}\mathrm{La}$ & -1.011 & -2.580 & -4.585 & 0.143  & 1.448  & -0.070 & 14.317 & 0.144  & 4.281  & 2.090  & 0.524         & -15.891 \\
		$^{209}_{\Lambda\Lambda}\mathrm{Pb}$ & 1.805  & 0.088  & -5.326 & 2.386  & 2.261  & 1.648  & 16.927 & 2.123  & 3.102  & 1.470  & 4.615         & -21.825 \\ 
		\hline\hline
	\end{tabular}
\end{sidewaystable}

\begin{sidewaystable}
	\centering
	\caption{RDF binding energies $E_{B}^{\rm RDF}(^{A+2}_{\Lambda\Lambda}Z)$ in MeV, directly calculated with each RDF functional.}
	\label{tab:EB_RDF}
	\renewcommand{\arraystretch}{1.05}
	\setlength{\tabcolsep}{7.1pt}
	\begin{tabular}{rrrrrrrrrrrrr}
		\hline\hline
		& PK1       & TM1       & NL-SH     & PKO1      & PKO2      & PKO3      & TW99      & PKDD      & DD-LZ1    & DD-ME2    & DD-ME$\delta$ & DD-MEX    \\
		$^{6}_{\Lambda\Lambda}\mathrm{He}$   & -33.786   & -35.219   & -34.267   & -32.215   & -31.662   & -30.894   & -29.410   & -31.924   & -30.414   & -31.716   & -34.728       & -31.755   \\
		$^{8}_{\Lambda\Lambda}\mathrm{Li}$   & -40.773   & -41.954   & -41.241   & -39.773   & -39.419   & -38.325   & -35.621   & -38.490   & -38.901   & -38.676   & -40.838       & -39.396   \\
		$^{9}_{\Lambda\Lambda}\mathrm{Li}$   & -49.725   & -50.616   & -50.306   & -48.699   & -48.523   & -47.290   & -44.220   & -47.200   & -48.532   & -47.360   & -49.257       & -48.384   \\
		$^{10}_{\Lambda\Lambda}\mathrm{Be}$  & -60.967   & -61.437   & -61.625   & -60.114   & -60.300   & -58.918   & -55.317   & -58.282   & -60.995   & -58.460   & -59.777       & -60.191   \\
		$^{11}_{\Lambda\Lambda}\mathrm{Be}$  & -71.074   & -70.851   & -71.797   & -69.961   & -70.226   & -68.978   & -65.490   & -68.256   & -72.399   & -68.474   & -69.095       & -70.718   \\
		$^{12}_{\Lambda\Lambda}\mathrm{B}$   & -82.254   & -81.108   & -82.963   & -81.097   & -81.507   & -80.367   & -76.719   & -79.279   & -85.107   & -79.509   & -79.015       & -82.604   \\
		$^{13}_{\Lambda\Lambda}\mathrm{B}$   & -98.651   & -96.269   & -99.363   & -96.883   & -97.322   & -96.513   & -93.564   & -95.544   & -103.428  & -95.815   & -94.282       & -99.659   \\
		$^{13}_{\Lambda\Lambda}\mathrm{C}$   & -95.545   & -93.306   & -96.310   & -94.356   & -94.764   & -93.923   & -90.327   & -92.431   & -100.091  & -92.680   & -91.181       & -96.482   \\
		$^{14}_{\Lambda\Lambda}\mathrm{C}$   & -115.752  & -111.830  & -116.303  & -113.643  & -114.022  & -113.575  & -110.890  & -112.387  & -122.340  & -112.547  & -109.535      & -117.408  \\
		$^{17}_{\Lambda\Lambda}\mathrm{N}$   & -142.461  & -143.828  & -142.706  & -141.375  & -140.381  & -140.825  & -138.275  & -141.337  & -143.666  & -141.424  & -144.319      & -144.229  \\
		$^{17}_{\Lambda\Lambda}\mathrm{O}$   & -138.745  & -140.162  & -139.049  & -138.189  & -137.150  & -137.607  & -134.382  & -137.537  & -139.909  & -137.634  & -140.457      & -140.448  \\
		$^{29}_{\Lambda\Lambda}\mathrm{Si}$  & -251.770  & -249.414  & -253.736  & -251.955  & -252.553  & -250.929  & -246.181  & -250.529  & -256.447  & -249.846  & -247.106      & -257.967  \\
		$^{33}_{\Lambda\Lambda}\mathrm{S}$   & -288.697  & -286.357  & -289.039  & -289.589  & -289.897  & -288.032  & -285.312  & -288.651  & -297.878  & -289.136  & -285.847      & -297.760  \\
		$^{41}_{\Lambda\Lambda}\mathrm{Ca}$  & -364.245  & -366.868  & -363.217  & -365.524  & -363.176  & -365.749  & -357.482  & -364.004  & -366.658  & -364.642  & -368.246      & -371.592  \\
		$^{52}_{\Lambda\Lambda}\mathrm{V}$   & -474.384  & -474.456  & -476.041  & -475.030  & -474.475  & -474.680  & -466.546  & -474.193  & -475.295  & -473.003  & -473.565      & -483.770  \\
		$^{90}_{\Lambda\Lambda}\mathrm{Y}$   & -810.878  & -812.872  & -812.096  & -812.289  & -809.916  & -813.036  & -799.288  & -811.137  & -810.625  & -809.988  & -812.051      & -825.669  \\
		$^{140}_{\Lambda\Lambda}\mathrm{La}$ & -1207.492 & -1210.308 & -1212.431 & -1206.446 & -1204.895 & -1206.994 & -1190.764 & -1206.139 & -1201.666 & -1204.440 & -1206.474     & -1225.039 \\
		$^{209}_{\Lambda\Lambda}\mathrm{Pb}$ & -1681.535 & -1683.957 & -1690.525 & -1681.146 & -1681.215 & -1682.361 & -1665.157 & -1681.138 & -1680.281 & -1682.472 & -1679.176     & -1709.100 \\
		\hline\hline
	\end{tabular}
\end{sidewaystable}
The evaluated double-$\Lambda$ binding energy deviations are obtained by inserting the single-$\Lambda$ deviations calculated with each RDF functional into the linear relation given in Eq.~\eqref{eq:LinearCorrelation}. The associated fitting uncertainty is $\sigma_{\rm Eval}=0.422$~MeV, and the resulting $\delta E_{B}^{\rm Eval}(^{A+2}_{\Lambda\Lambda}Z)$ values are listed in Table~\ref{tab:delta_eval}. The evaluated double-$\Lambda$ separation energies are then obtained using Eq.~\eqref{eq:BLambdaLambdaExp}. The RDF binding energies $E_{B}^{\rm RDF}(^{A+2}_{\Lambda\Lambda}Z)$ entering this expression are given in Table~\ref{tab:EB_RDF}, and the experimental core binding energies are taken from Ref. \cite{Wang2021CPC45.030003}. For each RDF functional, the evaluated double-$\Lambda$ separation energy $B_{\Lambda\Lambda}^{\rm Eval}$ carries the uncertainty $\sigma_{\rm Eval}$ from the linear-correlation analysis.

\section{C. Model selection and uncertainty prescription}\label{sec:selection}

To reduce the model dependence in the final evaluation, we select the RDF functionals by requiring consistency with the available double-$\Lambda$ data. For each RDF functional, the correlation-based procedure gives an evaluated double-$\Lambda$ separation energy $B_{\Lambda\Lambda}^{\rm Eval}$ with the uncertainty $\sigma_{\rm Eval}$. The functional is retained only if the evaluated interval overlaps with the corresponding experimental interval for all four measured double-$\Lambda$ hypernuclei, corresponding to the core nuclei $^{4}$He, $^{8}$Be, $^{9}$Be, and $^{11}$B~\cite{Ahn2013PRC88.014003, Ekawa2019PTEP2019.021D02, He2025Nat.Commun.16.11084}. For the $^{11}$B-core system, the larger experimental value, $B_{\Lambda\Lambda}=25.57\pm1.18$~MeV, is used in this selection~\cite{He2025Nat.Commun.16.11084}. With this criterion, the retained functionals are the density-dependent RHF functionals PKO1 and PKO2, and the density-dependent RMF functionals TW99, PKDD, DD-ME2, and DD-MEX.

For each double-$\Lambda$ hypernucleus, the final evaluated central value of $B_{\Lambda\Lambda}^{\rm Eval}$ is taken as the average over the retained RDF functionals,
\begin{equation}
	\mu=\frac{1}{n}\sum_{i=1}^{n}x_i,
\end{equation}
where $x_i$ denotes the evaluated separation energy obtained with the $i$th retained functional, and $n$ is the number of retained functionals. The model-to-model spread among the retained functionals is defined as
\begin{equation}
	\sigma_{\rm RDF}
	=
	\sqrt{
		\frac{1}{n}
		\sum_{i=1}^{n}
		(x_i-\mu)^2
	}.
\end{equation}
The uncertainty assigned to the final evaluated double-$\Lambda$ separation energy combines this spread with the uncertainty from the linear-correlation analysis,
\begin{equation}
	\sigma_{B_{\Lambda\Lambda}^{\rm Eval}}
	=
	\sqrt{\sigma_{\rm RDF}^{2}+\sigma_{\rm Eval}^{2}} .
\end{equation}
The evaluated $\Lambda\Lambda$ interaction energy is obtained from
\begin{equation}
	\Delta B_{\Lambda\Lambda}^{\rm Eval}
	=
	B_{\Lambda\Lambda}^{\rm Eval}
	-
	2B_{\Lambda}^{\rm Expt}.
\end{equation}
In propagating its uncertainty, the experimental uncertainty of $B_{\Lambda}^{\rm Expt}$ is assumed to be independent of $\sigma_{B_{\Lambda\Lambda}^{\rm Eval}}$, giving
\begin{equation}
	\sigma_{\Delta B_{\Lambda\Lambda}^{\rm Eval}}=
	\sqrt{\sigma_{B_{\Lambda\Lambda}^{\rm Eval}}^{2}+4\sigma_{B_{\Lambda}^{\rm Expt}}^{2}}.
\end{equation}

\end{document}